\documentclass[showpacs,amsmath,amssymb,twocolumn]{revtex4}

\usepackage{graphicx}
\usepackage{dcolumn}

\begin{document}

\title{Evidence of a pressure-induced metallization process in monoclinic VO$_2$}
\author{E. Arcangeletti$^1$, L. Baldassarre$^1$, D. Di Castro$^1$, S. Lupi $^1$, L. Malavasi$^2$, C. Marini$^1$, A. Perucchi$^1$, P. Postorino$^1$}
\affiliation{$^1$ ``Coherentia" CNR-INFM and Dipartimento di Fisica, Universit\`a di Roma La Sapienza,  Piazzale Aldo Moro 2, I-00185 Roma, Italy}
\affiliation{$^2$ Dipartimento di Chimica Fisica ``M. Rolla", INSTM and IENI-CNR, Universit\`a di Pavia, Viale Taramelli 16, I-27100 Pavia, Italy}

\begin{abstract}
Raman and combined trasmission and reflectivity mid infrared measurements have been carried out on  monoclinic
VO$_2$ at room temperature over the 0-19 GPa and 0-14 GPa pressure ranges, respectively. The pressure dependence
obtained for both lattice dynamics and optical gap shows a remarkable stability of the system up to P*$\sim$10
GPa. Evidence of subtle modifications of V ion arrangements within the monoclinic lattice together with the
onset of a metallization process via band gap filling are observed for P$>$P*. Differently from ambient
pressure, where the VO$_2$ metal phase is found only in conjunction with the rutile structure above 340 K, a new
room temperature metallic phase coupled to a monoclinic structure appears accessible in the high pressure
regime, thus opening to new important queries on the physics of VO$_2$.
\end{abstract}

\date{\today}

\maketitle

Since the first observation of the metal to insulator transition (MIT) in several vanadium oxides, these
materials attracted considerable interest because of the huge and abrupt change of the electrical properties at
the MIT. As usual in transition metal oxides, electronic correlation strongly affects the conduction regime of
vanadium oxides, although, in some compounds, lattice degrees of freedom seem to play an important role. This is
the case of VO$_2$, which undergoes a first order transition from a high temperature metallic rutile (R) phase
to a low temperature insulating monoclinic (M1) one. At the MIT temperature, $T_c$=340 K, the opening of an
optical gap in the mid-infrared (MIR) conductivity and a jump of several order of magnitude in the resistivity
are observed \cite{Morin}. The interest on this compound is thus mainly focused on understanding the role and
the relative importance of the electron-electron  and the electron-lattice interaction in driving the MIT.
Despite the great experimental and theoretical efforts \cite{tokura}, the understanding of this transition is
still far from being complete \cite{Kim,Biermann,Haverkor,Choi,Basov}. In the R phase the V atoms, each
surrounded by an oxygen octahedron, are equally spaced along linear chains in the \emph{c}-axis direction and
form a body-centered tetragonal lattice. On entering the M1 insulating phase the dimerization of the vanadium
atoms and the tilting of the pairs with respect to the \emph{c} axis
lead to a doubling of the unit cell, with space group changing from C$_{2h}^{5}$ (R) to
D$_{4h}^{14}$ (M1)  \cite{rutile,mono}. As first proposed by Goodenough \cite{Good}, the V-V pairing and the off-axis zig-zag
displacement of the dimers lead to a band splitting with the formation of a  Peierls-like gap at the Fermi level.
First principle electronic structure calculations
based on local density approximation (LDA) showed the band splitting on entering the monoclinic phase, but failed to yield the opening of the band gap
\cite{Wentz,Eyert}. In fact, as early pointed out \cite{ZylbeMott}, the electron-electron correlation has to be
taken into account to obtain the insulating phase. A recent theoretical paper where the electronic Coulomb
repulsion U is properly accounted for, shows that calculations carried out joining dynamical mean field theory
with the LDA scheme correctly captured both the rutile metallic and the monoclinic insulating state \cite{Biermann}.

There are many experimental evidences of strong electronic correlation in VO$_2$, as for example the anomalous
linear temperature dependence of conductivity above $T_c$ without any sign of saturation up to 800 K
\cite{resistivity}. Moreover a recent study of the temperature dependence of the optical conductivity
\cite{Basov} shows that MIT involves a redistribution of spectral weight (i.e the frequency-integrated optical
conductivity $\sigma_1$($\omega$)) within a broad energy scale ($\geqslant$ 5.5 eV). This suggests electronic
correlation to be crucial for the MIT, even though the insulating state may not be a conventional Mott insulator
because of the Peierls pairing. Finally it should be noted that insulating monoclinic VO$_2$ with a different
space group  C$_{2h}^{3}$ (M2) can be achieved under peculiar growing condition \cite{M2puro}, as well as by
means of minute amounts of Cr/V substitution \cite{M2}. In the M2 phase the Peierls pairing is partially
removed: one half of the V atoms dimerizes along the \emph{c}-axis, and the other one forms zig-zag chains of
equally spaced atoms \cite{M2}. The insulating character of the M2 phase has been reported as a further support
of the idea that the physics of VO$_2$ is close to that of a Mott-Hubbard insulator \cite{PougetStrain,comment}.

High pressure is an ideal tool for studying electron-correlated systems. Lattice compression in these systems
usually increases the orbital overlap and the electronic bandwidth (W) thus allowing for a systematic study of
the physical properties as a function of U/W. Unfortunately there is an almost complete lack of high pressure
experimental data on VO$_2$ apart from early resistivity measurements which show a small pressure-induced
increase (about 3 K) of  $T_c$ in the 0-4 GPa range \cite{PressJaya}. On the contrary the application of
pressure on Cr-doped VO$_2$ in the M2 phase induces a remarkable decrease (about 20 K) of $T_c$  within the 0-5
GPa range \cite{M2}.

In the present Letter we report on high-pressure MIR (0-14 GPa) and Raman (0-19 GPa) measurements  on VO$_2$ at
room temperature  to study pressure-induced effects on both the electronic structure and the lattice dynamics.
In particular, since the optical gap lies in the MIR region and the Raman spectra of VO$_2$ are drastically
different in the R and M1 phases, the techniques chosen allow to monitor independently the electronic and the
structural transitions. VO$_2$ was prepared starting from proper amounts of V$_2$O$_3$ and V$_2$O$_5$ (Aldrich
$>$99.9\%) pressed in form of pellet and reacted at 1050$^\circ$C in an argon flux for 12 h. Single crystal of
needle-like form were obtained. Phase purity was checked through single crystal x-ray diffraction on some
crystals and also on a powdered sample from crushed crystals. The room temperature refined lattice parameters
are in both cases in full agreement with the M1 monoclinic structure. The M1-R transition in synthesised samples
has been probed by means of Differential Scanning Calorimetry which revealed a clear endothermic peak at about
340 K in perfect agreement with the literature $T_c$ value.

A screw clamped opposing-plates diamond anvil cell (DAC) equipped with 400 $\mu$m culet II A diamonds  has been used for both Raman and MIR experiments. The gaskets were made of a 250 $\mu$m thick steel foil with a sample chamber of 130 $\mu$m diameter and 40 to 50 $\mu$m height under working conditions. We used NaCl an KBr as pressure transmitting media for Raman and MIR measurements respectively \cite{fit,PRB}. Pressure was measured \emph{in situ} with the standard
ruby fluorescence technique \cite{Mao}.

High-pressure MIR spectra of room temperature VO$_2$ have been collected exploiting the high brilliance of the
SISSI infrared beamline at ELETTRA synchrotron in Trieste \cite{sissi}. The incident and reflected (transmitted)
radiation were focused and collected by a cassegrain-based Hyperion 2000 infrared microscope equipped with a MCT
detector and coupled to a Bruker IFS 66v interferometer, which allows to explore the 750-6000 cm$^{-1}$ spectral
range. A VO$_2$ 5$\mu$m thick slab, obtained by pressing finely milled sample between the diamond anvils, have
been placed in the gasket hole, where a  KBr pellets was previously sintered \cite{PRLmanganiti}. The slits of
the microscope were carefully adjusted to collect transmitted and reflected light from the sample only and kept
fixed for all the experiment.

The high brilliance of the infrared synchrotron source and the proper sample thickness allow us to measure the
intensities reflected, $I_R^S$($\omega$), and transmitted, $I_T^S$($\omega$), at each pressure. Possible
misalignements and source intensity fluctuations have been accounted for by measuring the intensity reflected by
the external face of the diamond anvil, $I_R^{D}$($\omega$), at each working pressure. At the end of the pressure
run we measured the light intensities reflected by a gold mirror placed between the diamonds
$I_R^{Au}$($\omega$) and by the external face of the diamond anvil ${I_{R}^{D}}^{\prime}$($\omega$). By using the
ratio ${I_R^{D}}^{\prime}$($\omega$)/$I_{R}^{D}$($\omega$) as a correction function, we achieved the reflectivity
R($\omega$):
\begin{equation}
R(\omega)=\frac{I_R^S(\omega)}{I_R^{Au}(\omega)}\cdot\frac{{I_R^{D}}^{\prime}(\omega)}{I_{R}^{D}(\omega)}
\end{equation}
The  transmittance T($\omega$) is obtained using:
\begin{equation}
T(\omega)=\frac{I_T^S(\omega)}{I_T^{DAC}(\omega)}\cdot\frac{{I_R^{D}}^{\prime}(\omega)}{I_{R}^{D}(\omega)}
\end{equation}
where  $I_T^{DAC}(\omega)$ is the transmitted intensity of the empty DAC without gasket and with the anvils in
tight contact.
R($\omega$) and T($\omega$) at selected pressures are shown in Fig. 1.

\begin{figure}[b]
\includegraphics[width=6.4cm]{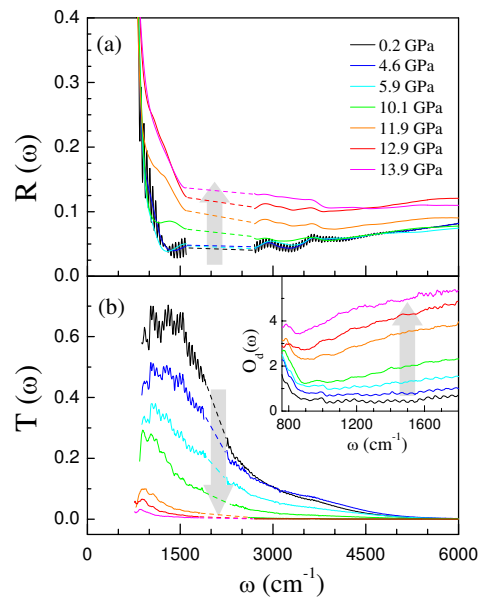}
\caption{(Color online). (a) MIR reflectivity R($\omega$) and (b) transmittance T($\omega$) of VO$_2$ at
selected pressures. Arrows indicate increasing pressure. Dashed lines are guides for the eyes which replaced the
data at the frequencies where the diamond absorption shows very intense peaks. For sake of clarity, fringes have
been averaged out in the reflectivity spectra, except at the lowest pressure. Inset: low-frequency optical
density $O_d(\omega)$.}
\end{figure}

At P $\sim$ 0, the low-frequency reflectivity  is characterized by a steep rise due to a phononic contribution,
whereas, on the high frequency side, R($\omega$) is slowly varying between 0.05 and 0.08, values slightly higher
than those expected for sample-diamond interface \cite{handbook,Barker2}. The extra reflectivity is due to the
multiple reflections within the sample-KBr bilayer which also originate the interference fringes well evident in
both R($\omega$) and T($\omega$). The low T($\omega$) values at high-frequency are due to the absorption of the
low-frequency tail of the electronic band \cite{Barker2, Basov}.

Within the 0-10 GPa pressure range R($\omega$) does not change remarkably whereas T($\omega$) gradually
decreases. On increasing pressure above 10 GPa R($\omega$) starts to increase and T($\omega$) abruptly decreases
to very small values owing to the shift of the electronic band towards lower frequencies. A clear evidence of
this process is found in the optical density $O_d(\omega)$= - $\ln{T(\omega)}$ shown in the inset of Fig. 1.
Above 10 GPa the electronic contribution fills the optical gap and the phonon peak is remarkably screened. These
findings show the onset of a pressure induced charge delocalization process. The simultaneous measurements of
R($\omega$) and T($\omega$) allow us to extract the real, n$(\omega)$, and the imaginary part, k$(\omega)$, of
the complex refractive index of VO$_2$. To this purpose a multilayer scheme diamond-sample-KBr-diamond has been
adopted, where multiple reflections within the sample and the KBr layers have been fully accounted for by adding
incoherently the intensities of the reflected beams \cite{Gruner}. Using the experimental layer thicknesses and
the known optical properties of diamond and KBr \cite{handbook}, R($\omega$) and T($\omega$) can be expressed as
a function of n$(\omega)$ and k$(\omega)$ only. The analytical derivation, under the above assumptions, is
straightforward albeit rather lengthy, and will be reported in detail in a forthcoming extended paper. The
non-analytical two equations system can be solved numerically  by an iterative procedure to obtain n$(\omega)$
and k$(\omega)$ \cite{iter}. The results obtained at the lowest pressure have been found to be in good agreement
with literature data at P=0 \cite{Basov,Barker2}. We point out that the independence of the results from the
guess function have been checked and that a complete agreement is found when a different calculation technique
\cite{Choi} is adopted to obtain n$(\omega)$ and k$(\omega)$.

The optical conductivity $\sigma_1(\omega)=2\omega/4\pi \cdot n(\omega)k(\omega)$ at different pressure is shown
in Fig. 2. The low frequency region of the spectra has not been reported because the strong variations of the
optical constants around the phonon contribution could affect the reliability of the results of the iterative
procedure. In any case the data shown in Fig. 2 allow to follow the pressure behaviour of the low frequency tail of
the electronic band, which is the spectral structure mostly affected by the MIT \cite{Basov}.

\begin{figure}
\includegraphics[width=7.1cm]{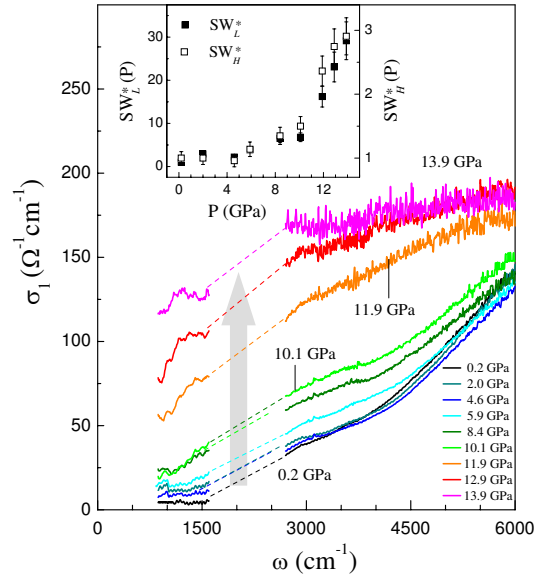}
\caption{(Color online). Optical conductivity $\sigma_1(\omega)$ of VO$_2$ at different pressures. Dashed lines
are a guide for the eyes. In the inset the normalized spectral weights calculated over the 900-1600 cm$^{-1}$
(SW$^*_{L}$(P)) and the 2600-5000 cm$^{-1}$ (SW$^*_{H}$(P)) frequency ranges are shown.}
\end{figure}

At P $\sim$ 0, $\sigma_1(\omega)$ is in good agreement with recent ambient pressure data \cite{Basov}
and it is weakly pressure dependent up to 4 GPa. On further increasing the pressure, $\sigma_1$($\omega$)
progressively increases, mainly within the 1500-4500 cm$^{-1}$ frequency range. Above 10 GPa an abrupt increase
of the overall $\sigma_1$($\omega$) occurs and a remarkable pressure-induced band gap filling is observed. The
data at the highest pressure clearly show that the energy gap, if still open, is well below 900 cm$^{-1}$. We
want to notice that a rough linear extrapolations of the data collected at P$>$10 GPa gives positive although small $\sigma_1$($\omega=0$) values compatible with a bad metal behaviour.

The effect of pressure can be better visualized if the pressure dependence of the spectral weight SW(P) is
analyzed. At each pressure we calculated the integral of $\sigma_1$($\omega$) over the 900-1600 cm$^{-1}$
[SW$_{L}$(P)] and 2600-5000 cm$^{-1}$ [SW$_{H}$(P)] frequency ranges. The integration has not been extended to
the frequency region above 5000 cm$^{-1}$, owing to the onset of saturation effects in the spectra collected at
the highest pressures (see in Fig. 2 the high noise level of these spectra at high frequencies). The spectral
weights normalized to the lowest pressure values, SW$^*_{L}$(P)=SW$_{L}$(P)/SW$_{L}$(0) and
SW$^*_{H}$(P)=SW$_{H}$(P)/SW$_{H}$(0) are shown in the inset of Fig. 2. Both the spectral weights show the same
pressure dependence, with a clear and abrupt change of slope at 10 GPa. We notice that the absolute
pressure-induced variation of SW$^*_{L}$(P) is much larger than that observed for SW$^*_{H}$(P), as expected if
charge delocalization occurs \cite{Basov}. Such a large and abrupt increase of the spectral weight in the gap
region is certainly compatible with the occurrence of a pressure induced MIT, although  the spectral range of
the present measurements does not allow to claim undoubtedly the complete optical gap closure above 10 GPa.

Raman measurements have been carried out using a confocal micro-Raman spectrometer equipped with a He-Ne laser
source (632.8 nm), a 1800 g/cm grating, and a Charge-Coupled-Device detector. A notch filter was used to reject
the elastic contribution of the backscattered light collected by a 20x objective. Under these experimental
conditions we achieved a  few microns diameter laser spot on the sample and a spectral resolution of about 3
cm$^{-1}$. Raman spectra of a VO$_2$ small single crystal collected within the 150-800 cm$^{-1}$ frequency range
are shown in Fig. 3 at selected pressures. The spectrum at the lowest pressure shows a full agreement with
previous data collected on VO$_2$ in the M1 phase at ambient pressure \cite{RamanOld,RamanNew}. Fifteen narrow
phonon peaks of the 18 Raman-active modes predicted for the M1 phase (9 A$_g$ + 9 B$_g$) can be identified in
the present measurements. The effect of pressure on the Raman spectrum results in a clear  phonon frequency
hardening, which, however, does not significantly changes either the peak pattern or the overall spectral shape
over the pressure range explored. Variations of the relative intensities of phonon peaks must be ascribed to
polarization effect, particularly strong in VO$_2$ \cite{RamanNew,Raman3}. Since the Raman spectrum of VO$_2$ in
the R phase is characterized by only 4 broad peaks \cite{RamanOld,RamanNew} we can safely conclude at a glance
that a transition to a R phase is not achieved by applying pressure up to 19 GPa. The good quality of the data
allows to apply a standard fitting procedure \cite{fit} to analyse the pressure dependence of the phonon
spectrum. The analysis shows an almost linear increase of the frequencies of every phonon peak eccept for
${\omega}_{V1}$ and ${\omega}_{V2}$ at 192 cm$^{-1}$ and 224 cm$^{-1}$ at P$\sim$0, respectively, whose pressure
dependence is shown in the inset of Fig. 3. A rather abrupt change of the  rate $d\omega/dP$ from a value close
to zero to 1-1.5 cm$^{-1}$/GPa is apparent at around 10 GPa.
\begin{figure}[t]
\includegraphics[width=6.8cm]{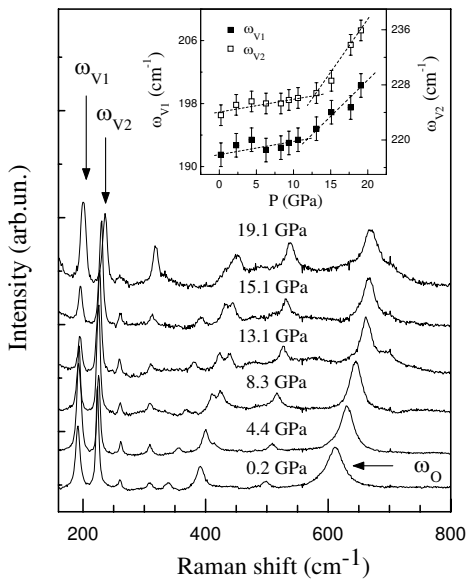}
\caption{Raman spectra of VO$_2$ at different pressures. Arrows mark the phonon peaks $\omega_{V1}$,  $\omega_{V2}$,
and $\omega_O$. Inset: phonon frequencies $\omega_{V1}$ and $\omega_{V2}$ as a function of pressure (dashed lines are guide for the eyes).}\label{raman}
\end{figure}
Due to the large difference between the V and the O mass, the low frequency  $\omega_{V1}$ and $\omega_{V2}$
peaks can be ascribed to the V-ions motion in the dimerized chains. In fact by comparing the phonon frequencies
of VO$_2$ with those of NbO$_2$, which has the same crystal structure and exhibits a quite similar Raman
spectrum \cite{NbO2}, $\omega_{V1}$ and $\omega_{V2}$ frequencies scale with the mass of the transition metal,
while the frequency of the $\omega_O$ peak (see Fig. 3) scales with the reduced mass between the oxygen and the
transition metal. Albeit the monoclinic structure is retained over the whole pressure range explored, the latter
finding suggests some rearrangement of the V-V chains as the pressure is increased above 10 GPa.

To summarize, we report on a careful high-pressure Raman and MIR investigation of VO$_2$. In particular, taking
full advantage from the high brilliance of the SISSI beamline, we were able to collect both transmittance and
reflectivity of the sample in the DAC allowing for a full analysis of the MIR data. The whole of the results
clearly identifies two regimes below and above a threshold pressure P*$\sim$ 10 GPa. Both Raman and IR spectra
show a weak pressure dependence for P$<$P* whereas, for P$>$P*, the pressure driven optical gap filling is accompanied
by a rearrangement of the V chains within a monoclinic framework. The stability of the monoclinic phase,
which retains up to 19 GPa, and the pressure-driven delocalization process, which starts at P* and it is quite
remarkable at the highest pressures (14 GPa), indicate that the metal-insulator and monoclinic-rutile transitions are decoupled in high pressure VO$_2$. Recent high-pressure x-ray diffraction data \cite{Malavasi} which extend the stability
of the monoclinic phase up to 42 GPa further support the above finding.

The rearrangement of the V chains at P* suggests the occurrence of a subtle transformation into a new monoclinic
phase where lattice compression enables the system to evolve towards a metallic phase.  A pressure-induced
transition from the M1 to the M2 phase can be conjectured, bearing in mind the strong pressure-induced reduction
of the MIT temperature observed in M2 Cr-doped  VO$_2$ \cite{M2}. This transition is compatible with our Raman
results since the spectra of VO$_2$ in the M1 and M2 phases are expected quite similar (group theory predicts
the same number of Raman modes). The present results thus open to new experimental queries and represent a
severe benchmark for theoretical model aimed at addressing the role of the electron-electron correlation and the
structural transition in driving the MIT in VO$_2$.


\begin{thebibliography}{99}
\bibitem{iter}The guess function n$^0$($\omega$) was put equal to the average refraction index  of room
temperature VO$_2$ \cite{Barker2} and k$^0$($\omega$) is obtained inserting
n$^0$($\omega$) into the T($\omega$) equation. The successive iterate, n$^1$($\omega$), is obtained inserting
k$^0$($\omega$) into R($\omega$) and in turn n$^1$($\omega$) into T($\omega$) to get k$^1$($\omega$). The
procedure has been iterated till convergence.
\end{thebibliography}
\end{document}